\documentclass{JHEP3}
\usepackage{graphicx,amsmath,amssymb}
\usepackage{epsfig,multicol}
\usepackage{epsf}
\usepackage{epstopdf}
\input{epsf.sty}

\usepackage{graphicx}
\usepackage{amssymb}
\usepackage{amsmath}

\def\p{\partial}

\def\half{{1\over 2}}
\def\({\left(}
\def\){\right)}
\def\[{\left[}
\def\]{\right]}

\def\e{\begin{equation}}
\def\q{\end{equation}}
\def\m{\begin{eqnarray}}
\def\n{\end{eqnarray}}

\title{The Trispectrum in the Multi-brid Inflation}
\author{Qing-Guo Huang \footnote{huangqg@kias.re.kr}
\\\small{\em School of Physics, Korea Institute for Advanced Study,
207-43, Cheongryangri-Dong, Dongdaemun-Gu, Seoul 130-722, Korea } }

\abstract{ The trispectrum is at least as important as the bispectrum and its size can be characterized by two parameters $\tau_{NL}$ and $g_{NL}$. In this short paper, we focus on the Multi-brid inflation, in particular the two-brid inflation model in arXiv.0805.0974, and find that $\tau_{NL}$ is always positive and roughly equals to $({6\over 5}f_{NL})^2$ for the low scale inflation, but $g_{NL}$ can be negative or positive and its order of magnitude can be the same as that of $\tau_{NL}$ or even larger. }

\keywords{Trispectrum, Multi-brid inflation}

\begin{document}

\section{Introduction}

The single-field slow-roll inflation predicts a roughly Gaussian distribution of the primordial power spectrum \cite{Maldacena:2002vr}. However in the general case with a large number of fields, a large deviation from Gaussian distribution is expected. A well-understood ansatz of non-Gaussianity has a local shape. Working in the framework of Fourier
transformation of $\zeta$, the primordial power spectrum ${\cal
P}_\zeta$ is defined by 
\e 
\langle\zeta({\bf k_1})\zeta({\bf
k_2})\rangle=(2\pi)^3 {\cal P}_{\zeta}(k_1)\delta^3({\bf k_1}+{\bf
k_2}), 
\q  
and the primordial bispectrum and trispectrum are defined by 
\m
\langle\zeta({\bf k_1})\zeta({\bf k_2})\zeta({\bf
k_3})\rangle&=&(2\pi)^3 B_\zeta(k_1,k_2,k_3)\delta^3({\bf k_1}+{\bf k_2}+{\bf k_3}), \\
\langle\zeta({\bf k_1})\zeta({\bf k_2})\zeta({\bf k_3})\zeta({\bf
k_4})\rangle&=&(2\pi)^3 T_\zeta(k_1,k_2,k_3,k_4)\delta^3({\bf
k_1}+{\bf k_2}+{\bf k_3}+{\bf k_4}). 
\n 
The bispectrum and
trispectrum are respectively related to the power spectrum by 
\m
B_\zeta(k_1,k_2,k_3)&=&{6\over 5}
f_{NL}[{\cal P}_\zeta(k_1){\cal P}_\zeta(k_2)+2\ \hbox{perms}], \label{bi} \\
T_\zeta(k_1,k_2,k_3,k_4)&=&\tau_{NL}[{\cal P}_\zeta(k_{13}){\cal
P}_\zeta(k_3){\cal P}_\zeta(k_4)+11\ \hbox{perms}] \nonumber \\
&+&{54\over 25}g_{NL}[{\cal P}_\zeta(k_2){\cal P}_\zeta(k_3){\cal
P}_\zeta(k_4)+3\ \hbox{perms}], \label{tri} 
\n 
where $f_{NL}$, $\tau_{NL}$ and $g_{NL}$ are the non-Gaussianity parameters which measure the size of the non-Gaussianity.

In the $\delta N$ formalism \cite{Starobinsky:1986fxa}, the curvature perturbation in the multi-field inflation model can be expanded as
\m
\zeta&=&\delta N=N(\phi_i+\delta\phi_i)-N(\phi_i) \nonumber \\ 
&=&N_{,i}\delta\phi_i+\half N_{,ij}\delta\phi_i\delta\phi_j+{1\over 6}N_{,ijk}\delta\phi_i\delta\phi_j\delta\phi_k +...\ , 
\n
where
\e
N_{,i}={\p N\over \p\phi_i},\quad N_{,ij}={\p^2 N\over \p\phi_i\p\phi_j},\quad N_{,ijk}={\p^3 N\over \p\phi_i\p\phi_j\p\phi_k},
\q
and the repeated sub-indices are summed over. Therefore the non-Gaussianity parameters are given by, \cite{Alabidi:2005qi},
\m
f_{NL}&=&{5\over 6}{N_{,ij}N_{,i}N_{,j}\over (N_{,l}N_{,l})^2},\\
\tau_{NL}&=&{N_{,ij}N_{,ik}N_{,j}N_{,k}\over (N_{,l}N_{,l})^3},\\
g_{NL}&=&{25\over 54} {N_{,ijk}N_{,i}N_{,j}N_{,k}\over (N_{,l}N_{,l})^3}.
\n
From the Cauchy-Schwarz inequality, the $\tau_{NL}$ has a lower bound,  \cite{Suyama:2007bg},
\e
\tau_{NL}\geq ({6\over 5}f_{NL})^2.
\q
The above inequality is saturated in the single field case, or the vector $N_{,i}$ is an eigenvector of the matrix $N_{,ij}$ in the case with multi fields. So $\tau_{NL}$ is expected to be large if $f_{NL}\gg 1$. Since $N_{,ijk}$ is quite model-dependent, $g_{NL}$ can be negative or positive, and its order of magnitude can be large or small.

WMAP 5yr data \cite{Komatsu:2008hk} implies 
\e
-9<f_{NL}^{local}<111
\q
at $2\sigma$ level. Even though the Gaussian distribution is still consistent with data, the allowed negative part of $f_{NL}^{local}$ has been cut from the WMAP 3yr data significantly. For example, if $\tau_{NL}>560$, it can also be detected by Planck \cite{Kogo:2006kh}. In \cite{Kogo:2006kh} the authors also pointed out that the trispectrum of the Planck data is more sensitive to primordial non-Gaussianity than the bispectrum for $|f_{NL}|\gtrsim 50$. The trispectrum is at least as important as the bispectrum.

As we know,  a large local-shape $f_{NL}$ can be achieved in the multi-brid inflation. In this short paper, we investigate the trispectrum  in the special two-brid inflation model in \cite{Sasaki:2008uc}. One can discuss more complicated setups, such as \cite{Naruko:2008sq}. However, we will see that the phenomenology in this simple model is rich enough. In Sec.2, we calculate the trispectrum in this two-brid inflation model and discuss its property in detail. Some discussions are given in Sec. 3.

\section{The trispectrum in the Multi-brid inflation model}

In this section, we consider the two-brid inflation in \cite{Sasaki:2008uc} with the potential 
\e 
V(\phi_1,\phi_2)=V_0 \exp\(\alpha_1\phi_1+\alpha_2\phi_2\),
\q
where $\alpha_1$ and $\alpha_2$ are two free parameters in the unit of $M_p=1$. The effective mass for each inflaton is $m_i=\sqrt{3}\alpha_iH_*$ where $H_*$ is the Hubble parameter during inflation. Here the inflatons are assumed to slowly roll down their potentials and the slow-roll conditions require $\alpha_i\ll 1$ for $i=1,2$. The dynamics of these two inflatons is governed by 
\m
{d\phi_i\over dN}\simeq \alpha_i, \quad (i=1,2),
\n
where $dN=-Hdt$.  In \cite{Sasaki:2008uc}, $V_0$ is prompted to be
\e
V_0=\half \sum_{i=1}^2g_i^2\phi_i^2\chi^2+{\lambda\over 4}(\chi^2-{\sigma^2\over \lambda})^2.
\q
Therefore the inflation ends when $\phi_i=\phi_{i,f}$ which is given by
\e
\sum_{i=1}^2 g_i^2\phi_{i,f}^2=\sigma^2.
\q
For convenience, we parametrize $\phi_{i,f}$ as
\e
\phi_{1,f}={\sigma\over g_1}\cos\gamma,\quad \phi_{2,f}={\sigma\over g_2}\sin\gamma.
\q
Without loss of generality, we assume $\alpha_i\phi_i>0$ (for $i=1,2$).

The number of e-folds before the end of inflation is given by
\e
N=\half\ln \({e^{2\phi_1/\alpha_1}+e^{2\phi_2/\alpha_2} \over  e^{2\phi_{1,f}/\alpha_1} +e^{2\phi_{2,f}/\alpha_2 } }\).
\q
Because the surface where the inflation ends in the field space is not exactly the surface of constant energy density, the author in \cite{Sasaki:2008uc} introduced a correction to the number of e-folds before the end of inflation as
\e
N_c={\sigma\over 4}({\alpha_1\over g_1}\cos\gamma+{\alpha_2\over g_2}\sin\gamma).
\q 
However, this correction term can be neglected if $\alpha_i\ll 1$. So we ignore it in our paper.
In order to simplify the notation, we define some new parameters as
\m
g&=&\sqrt{g_1^2\cos^2\gamma+g_2^2\sin^2\gamma}, \label{defg}\\
\alpha_1&=&\alpha\cos\theta, \quad \alpha_2=\alpha\sin\theta.
\n
We expand $\delta N$ to the third order,
\m 
\delta N&=&{g_1\cos\gamma\delta\phi_1+g_2\sin\gamma\delta\phi_2 \over \alpha g c_+} + {g_1^2g_2^2\over 2\sigma g^3} {(\sin\theta\delta\phi_1-\cos\theta\delta\phi_2)^2 \over \alpha c_+^3} \nonumber \\ &-&{g_1^3g_2^3\over 2\sigma^2 g^4}{c_-(\sin\theta\delta\phi_1-\cos\theta\delta\phi_2)^3\over \alpha c_+^5}, 
\n 
where
\m
c_-&=&{g_1\over g}\cos\theta\sin\gamma-{g_2\over g}\sin\theta\cos\gamma,\\
c_+&=&{g_1\over g}\cos\theta\cos\gamma+{g_2\over g}\sin\theta\sin\gamma.
\n

Assume that the scalar field fluctuations $\delta\phi_1$ and $\delta\phi_2$ are Gaussian and non-correlated, 
\e
\langle \delta\phi_i \delta\phi_j \rangle=\({H_*\over 2\pi}\)^2\delta_{ij},
\q
where $H_*$ denotes the Hubble parameter at the time of Hubble exit. Hence the amplitude of the primordial power spectrum is 
\e 
P_\zeta={1\over \alpha^2 c_+^2}\({H_*\over 2\pi}\)^2.
\q
Since the amplitude of the tensor perturbation is still given by
\e
P_T=8\({H_*\over 2\pi}\)^2,
\q
the tensor-scalar ratio becomes 
\e
r={P_T/P_\zeta}=8\alpha^2 c_+^2.
\q
We can also easily calculate the spectral index, namely
\e
n_s=1-\alpha^2.
\q
The spectral index and tensor scalar ratio can be measured by the experiments. Once they are fixed, the parameters $\alpha$ and $c_+$ in this two-brid inflation model are fixed as well. 
WMAP 5yr data \cite{Komatsu:2008hk} implies 
\e
n_s=0.96_{-0.013}^{+0.014}. 
\q
For $n_s=0.96$, $\alpha=0.2$. On the other hand, we can prove that $c_+^2\leq 1$ which implies
\e
r\leq 8(1-n_s), \label{cons}
\q
where the equality is satisfied when 
\e
g_1\sin\theta\cos\gamma=g_2\cos\theta\sin\gamma.
\q
Eq.(\ref{cons}) can be taken as a consistency relation for this two-brid inflation model. For $n_s=0.96$, $r\leq 0.32$ which is consistent with WMAP 5yr data $(r<0.20)$.

From Eqs.(\ref{bi}) and (\ref{tri}), the non-Gaussiniaty parameters are given by
\m
f_{NL}&=&{5\alpha g_1^2g_2^2\over 6\sigma g^3}{{\tilde c}^2\over c_+},\\
\tau_{NL}&=&{\alpha^2 g_1^4g_2^4\over \sigma^2 g^6}{{\tilde c}^2\over c_+^2},\\
g_{NL}&=&{25\alpha^2 g_1^3g_2^3\over 18\sigma^2 g^4}{c_- {\tilde c}^3\over c_+^2},
\n
where
\e
{\tilde c}={g_2\over g}\cos\theta\sin\gamma-{g_1\over g}\sin\theta\cos\gamma.
\q
We  see that the second order non-Gaussianity parameters are related to $f_{NL}$ by
\m
\tau_{NL}&=&{36\over 25}{1\over {\tilde c}^2}f_{NL}^2,\\
g_{NL}&=&{2g^2\over g_1g_2}{c_-\over {\tilde c}}f_{NL}^2.
\n
Since 
\e
{\tilde c}^2=1-c_+^2
\q 
and $c_+^2\leq 1$, ${\tilde c}^2\leq 1$ and thus 
\e
\tau_{NL}\geq ({6\over 5}f_{NL})^2.
\q
The inequality is saturated when 
\e
g_1\cos\theta\cos\gamma=-g_2\sin\theta\sin\gamma.
\q
Keeping $f_{NL}$ fixed, $\tau_{NL}$ can be much larger than $f_{NL}^2$ if $c_+\rightarrow 1$. However, in the limit of $c_+\rightarrow 1$, $f_{NL}$ should approach to 0.
Here we want to remind the readers that both $\alpha$ and $c_+$ can be fixed by the spectral index and the tensor-scalar ratio.  Considering 
\e
c_+^2={r\over 8(1-n_s)},
\q
$c_+^2<0.625$ for $n_s=0.96$ and $r<0.20$. If $r\ll 1$, $c_+\sim 0$ and then $\tau_{NL}\simeq {36\over 25}f_{NL}^2$.

\subsection{$g_1=g_2$}

From Eq.(\ref{defg}), $g_1=g_2=g$ in this case.  Now $c_-$, ${\tilde c}$, and $c_+$ are simplified to be
\e
c_-={\tilde c}=\sin(\gamma-\theta),\quad c_+=\cos(\gamma-\theta).
\q
Therefore 
\m
f_{NL}&=&{5\alpha g\over 6\sigma} ({1\over c_+}-c_+),\\
\tau_{NL}&=&{36\over 25(1-c_+^2)}f_{NL}^2,\\
g_{NL}&=&2f_{NL}^2.
\n
In this special case, $g_{NL}$ is positive and is related to $f_{NL}$ by $2f_{NL}^2$.
If $r\ll 1$, $c_+\sim 0$ and then $\tau_{NL}\simeq {36\over 25}f_{NL}^2$ which is roughly the same order of magnitude as $g_{NL}$, and 
\e
f_{NL}\simeq {5\alpha g\over 6\sigma c_+}.
\q

Because $\chi=0$ during inflation, the Hubble parameter is related to $\sigma$ by  
\e
H_*^2={\sigma^4\over 12\lambda}. 
\q
Considering $H_*=1.1\times 10^{-4}r^\half$, $\sigma$ is related to $r$ by
\e
\sigma=1.95\times 10^{-2}\lambda^{1\over 4}r^{1\over 4}.
\q
If $r\ll1$, the non-Gaussianity parameter $f_{NL}$ takes the form 
\e
f_{NL}\simeq 121(1-n_s){g\over r}({r\over \lambda})^{1\over 4}.
\q
On the other hand, $\sigma$ is required to be large compared to the Hubble parameter during inflation for the field $\chi$ to work as a water-fall field. So we have 
\e
(r/\lambda)^{1\over 4}\lesssim 177.
\q
Therefore 
\e
f_{NL}\lesssim 2.1\times 10^4 (1-n_s) {g\over r}. 
\q
For $n_s=0.96$, $f_{NL}\lesssim 810 g/r$.  

To summarize, if $g_1=g_2$, $f_{NL}$ is positive and can be very large, and $g_{NL}=2f_{NL}^2$. In this case, $\tau_{NL}\geq {36\over 25}f_{NL}^2$ and the equality is roughly satisfied if $r\ll 1$.

\subsection{$g_1\neq g_2$}

If $g_1=g_2$, we find that $g_{NL}=2f_{NL}^2$ which must be positive. In this subsection, we focus on the case of $g_1\neq g_2$ and invistigate whether $g_{NL}$ can be negative, but the absolute value is still large.
Since there is a symmetry between $\phi_1$ and $\phi_2$, we assume 
\e
\Delta={g_1\over g_2}< 1,
\q 
without loss of generality.

Keeping $f_{NL}$ fixed, a large absolute value of $g_{NL}$ might be obtained if  ${\tilde c}=0$ which implies
\e
\gamma=\gamma_0=\tan^{-1}(\Delta \tan\theta).
\q
On the other hand, $f_{NL}$ should be 0 when ${\tilde c}=0$. So we consider that $\gamma$ slightly deviates from $\gamma_0$, namely
\e
\gamma=\gamma_0+\delta,
\q
with $\delta\ll \gamma_0$. For $\Delta\ll 1$, we have
\m
g&\simeq&{g_1\over \kappa},\\
{\tilde c}&\simeq&{\kappa^2\over \Delta}\delta,\\
c_-&\simeq&-{\sin 2\theta\over 2\Delta},\\
c_+&\simeq&1,
\n
where
\m
\kappa=\sqrt{\cos^2\theta+\Delta^2\sin^2\theta}.
\n
In this limit, $c_+\simeq 1$ and then the tensor-scalar ratio is slightly larger than the WMAP bound if $n_s=0.96$. Here we ignore this constrant first and illustrate an important observation. Now we have 
\e
f_{NL}\simeq {5\alpha g_2\over 6\sigma}{\kappa^7\over \Delta^3}\delta^2,
\q
and 
\m
\tau_{NL}&\simeq&{36\over 25}{\Delta^2\over \kappa^4\delta^2}f_{NL}^2,\\
g_{NL}&\simeq&-{\Delta\sin 2\theta\over \kappa^4\delta}f_{NL}^2.
\n
Since both $\Delta$ and $\sin 2\theta$ are positive, $g_{NL}$ can be negative if $\delta>0$.  In order to get the precise results and go beyond the above approximation, we should adopt the numerical calculation. For example, inputting $n_s=0.96$, $r=0.22$, $\Delta=0.1$, we have 
\e
\tau_{NL}/f_{NL}^2\simeq 4.6,
\q
and 
\m
\theta=0.79, \ \gamma=0.466 \ &\rightarrow& \ f_{NL}=0.03{\alpha g_2\over \sigma},\ g_{NL}/f_{NL}^2\simeq -10;\\
 \theta=1.38, \ \gamma=0.1 \ &\rightarrow& \ f_{NL}=1.12{\alpha g_2\over \sigma},\ g_{NL}/f_{NL}^2\simeq 4.92.
\n
So there is still a big room for a large $f_{NL}$. We see that a negative $g_{NL}$ with large absolute value can be obtained in the case of $g_1\neq g_2$.

\section{Conclusions}

The non-Gaussianity is expected to be large in the multi-brid inflation. In the simple two-brid inflation model \cite{Sasaki:2008uc}, $\tau_{NL}$ must be positive and roughly equals to $({6\over 5}f_{NL})^2$ due to the upper bound on the inflation scale $(r<0.2)$, but $g_{NL}$ can be negative or positive and its order of magnitude can be the same as that of $\tau_{NL}$ or even larger. In \cite{Byrnes:2008zy} the authors consider a more complicated case where $\tau_{NL}$ can be much larger than $({6\over 5}f_{NL})^2$, but it seems unlikely for the case with red-tilted power spectrum and large positive $f_{NL}$.

As we know, multi-brid inflation model and curvaton model are the only cases that can be analyzed
systematically (over a fairly wide range of the model parameters) and are capable of yielding fairly
large local-shape $f_{NL}^{local}$. In the usual single curvaton model, the curvature perturbation is assumed to be generated by curvaton and then we have $\tau_{NL}=({6\over 5}f_{NL})^2$. The size of $g_{NL}$ depends on the curvaton potential. If the relevant part of curvaton potential takes the exactly quadratic form, we have $g_{NL}=-{10\over 3}f_{NL}$ .  If a non-quadratic correction becomes visible compared to the quadratic term, $g_{NL}$ in the curvaton model can has a large deviation from the relation $g_{NL}=-{10\over 3}f_{NL}$ and the sign of $g_{NL}$ depends on the sign of the non-quadratic term which can be positive or negative as well. The order of magnitude of $g_{NL}$ for the curvaton model with non-quadratic correction can be roughly the same as $f_{NL}^2$ which takes the same order of $\tau_{NL}$. See \cite{Sasaki:2006kq} in detail. If the curvaton potential is dominated by the non-quadratic term, $\tau_{NL}\sim g_{NL}\sim f_{NL}^2$ \cite{Huang:2008zj}. On the other hand, in the mixed curvaton scenario \cite{Huang:2008zj,Huang:2008rj} where the total curvature perturbation is still mainly produced by the inflaton, not curvaton, $\tau_{NL}$ is enhanced by a large factor $1/\beta$ with respect to $({6\over 5}f_{NL})^2$, where $\beta$ measures the size of curvature perturbation caused by curvaton compared to the total curvature perturbation including the contribution from inflaton. So $\tau_{NL}$ can be much larger than $({6\over 5}f_{NL})^2$ in the mixed curvaton scenario.  This is different from the multi-brid inflation. In the mixed curvaton model, whether $g_{NL}$ is enhanced or suppressed depends on how $g_{NL}^{cv}$ is related to $f_{NL}^{cv}$, where the superscript ``cv" denotes the non-Gaussianity generated by curvaton in the usual curvaton model. For example, if $g_{NL}^{cv}\sim f_{NL}^{cv}$, $g_{NL}\sim \beta f_{NL}$ which is expected to be very small and undetectable, but if $g_{NL}^{cv}\sim (f_{NL}^{cv})^2$, $g_{NL}\sim (f_{NL})^2/\beta$ which is enhanced and may be detectable. If we can not only detect $f_{NL}$, but also get some restricted constraints on $\tau_{NL}$ and $g_{NL}$ in the forthcoming experiments, it is possible to distinguish multi-brid inflation model from curvaton model.

Recently there are many papers concerning on the brispectrum and the trispectrum \cite{Huang:2008ze,Byrnes:2007tm}.  Here we want to stress that the trispectrum might be as important as the bispectrum and we encourage more theorists and experimenters  to pay more attentions to the trispectrum in the near future.

\vspace{1.4cm}

\noindent {\bf Acknowledgments}

\vspace{.5cm}

We would like to thank M.~ Sasaki for very helpful discussions.

\end{document}